\documentclass[a4paper]{article}

\usepackage{INTERSPEECH2020}

\usepackage{color, hyperref, verbatim}
\hypersetup{
    colorlinks,
    citecolor=black,
    filecolor=black,
    linkcolor=black,
    urlcolor=black
}
\usepackage{multirow}

\title{RECOApy: Data recording, pre-processing and phonetic transcription for end-to-end speech-based applications}
\name{Adriana STAN}

\address{
Communications Department \\Technical University of Cluj-Napoca, Romania}
\email{adriana.stan@com.utcluj.ro}

\begin{document}

\maketitle
\begin{abstract}
Deep learning enables the development of efficient end-to-end speech processing applications while bypassing the need for expert linguistic and signal processing features. Yet, recent studies show that good quality speech resources and phonetic transcription of the training data can enhance the results of these applications. In this paper, the RECOApy tool is introduced. RECOApy streamlines the steps of data recording and pre-processing required in end-to-end speech-based applications. The tool implements an easy-to-use interface for prompted speech recording, spectrogram and waveform analysis, utterance-level normalisation and silence trimming, as well grapheme-to-phoneme conversion of the prompts in eight languages: Czech, English, French, German, Italian, Polish, Romanian and Spanish.

The grapheme-to-phoneme (G2P) converters are deep neural network (DNN) based architectures trained on 
 lexicons extracted from the Wiktionary online collaborative resource. With the different degree of orthographic transparency, as well as the varying amount of phonetic entries across the languages, the DNN's hyperparameters are optimised with an evolution strategy. The phoneme and word error rates of the resulting G2P converters are presented and discussed. The tool, the processed phonetic lexicons and trained G2P models are made freely available. 

\end{abstract}

%%%%%%%%%%%%%%%%%%%%%%%%
%% Introduction
%%%%%%%%%%%%%%%%%%%%%%%%
\section{Introduction}
\label{sec-intro}

Nowadays, in the development of deep neural networks (DNN) based speech processing applications, most of the signal preprocessing, feature extraction and linguistic annotations are part of the inherent neural learning. This means that systems for automatic speech recognition (ASR) and text-to-speech synthesis (TTS) can be easily trained using only pairs of audio and orthographic transcript \cite{DeepSpeech,Tacotron2,DeepVoice}. A major advantage of this approach is that training data can be easily and readily found, and that there is no language dependency in the development stage---other than the language specific speech resources. 
Although this approach yields satisfactory results for most end-user applications, when it comes to high quality systems, found speech data and orthographic input does not suffice \cite{fong2019}. Most of the high-end commercial applications still make use of large amounts of studio recordings and elaborate text processing modules \cite{Tacotron2,48239}.

Hence, there is still a need for tools which can facilitate the development of domain or speaker specific training data, as well as tools which can generate expert linguistic features in a variety of languages. In this context, the first version of the RECOApy tool is introduced. RECOApy was designed with the main purpose of enabling end-users to record their own data and prepare it for end-to-end speech processing applications. It provides an easy to use interface for prompted speech recoding which includes several monitoring and data processing options (see Section \ref{sec-reco}), as well as a set of highly accurate pre-trained neural network models able to phonetically transcribe the prompts in eight languages.

The task of building grapheme-to-phoneme converters is not novel, but
depending on a language's orthographic transparency and onset entropy \cite{gillon}, G2P can be solved using simple rule-based systems (e.g. Finnish) or can pose serious problems even for the most advanced deep learning algorithms (e.g. English). The modern G2P converters aim at solving the problem of phonetic transcription in multiple languages at once. But phonetic lexicons are not readily available in most languages, and researchers are now investigating the use of collaborative online resources, such as Wiktionary,\footnote{\url{www.wiktionary.org}} as an alternative. \cite{schlippe2014web}~does just this by extracting the phonetic transcriptions in six languages from Wiktionary and validates them over manually crafted lexicons. The authors of \cite{deri-knight:2016:ACL} also use several online repositories to train and adapt the models from high-resource languages to related low-resource languages. 
Multilingual G2P was also addressed by changing the grapheme representation: \cite{mingzhi2020} proposes a model which uses byte-level input representation to accommodate different grapheme systems, along with an attention-based Transformer architecture. Ancillary audio data was also used to learn a more optimal intermediate representation of source graphemes in a multi-task training process for multilingual G2P~\cite{route-etal-2019-multimodal}. 
%Their architecture yields 16.2\%– 50.2\% relative word error rate improvement over character-based counterparts for mono- and multi-lingual use cases.

As the grapheme-to-phoneme task is inherently a sequence to sequence (\emph{seq2seq}) mapping problem, the G2P converter in RECOApy uses this type of learning architecture. Similar approaches were introduced in \cite{Yao2015SequencetosequenceNN}. The authors map the entire input grapheme sequence to a vector, and then use a recurrent neural network to generate the output sequence conditioned on the encoding vector. \cite{43264} describes a G2P model based on a unidirectional LSTM with different output delays and deep bidirectional LSTM with a connectionist temporal classification layer. 
Milde \emph{et al.}~\cite{Milde2017MultitaskSM} investigate how multitask learning can improve the performance of sequence-to-sequence G2P models. A single seq2seq model is trained on multiple phoneme lexicon datasets containing several languages and phonetic alphabets. Esch \emph{et al.}~\cite{45913} train recurrent neural network-based models to predict the syllabification and stress patterns of the input text for TTS, while also deriving phonetic transcriptions in the process.
The use of entire phrases as input to LSTM, biLSTM and CNN-based neural networks and their evaluation in English, Czech and Russian is presented in \cite{Juzov2019UnifiedLD}.

Starting from this overview of multilingual and neural networks-based training schemes,
RECOApy's G2P module incorporates the use of online collaborative phonetic lexicons and lexicon-tailored seq2seq neural network architectures derived with the help of an evolution strategy. The RECOApy tool, along with the parsed lexicons and complete set of trained models are made freely available. The G2P module can be used as a standalone tool as well. 

The paper is organised as follows: Section \ref{sec-reco} introduces the recording app and its features. Section~\ref{sec-phon} presents the phonetic transcription tool development and hyperparameter tuning using evolution strategies. Results of the phonetic converters are discussed in Section~\ref{sec-res}, and conclusions are drawn in Section~\ref{sec-conc}. 

%%%%%%%%%%%%%%%%%%%%%%%%
%% RecoApPy
%%%%%%%%%%%%%%%%%%%%%%%%
\vspace{-0.2cm}
\section{RECOApy GUI}
\label{sec-reco}

Recording prompted speech by end-users can be easily performed with any of the numerous 
free general purpose recording tools available, such as Audacity\footnote{\url{https://www.audacityteam.org/}} or Wavesurfer \cite{conf/interspeech/SjolanderB00}. But this means that in order to obtain phrase-length speech segments, the continuous recording stream needs to be manually segmented and aligned to the prompts. Or that the recording operator needs to start and stop the recording after each prompt reading. In both cases incorrect readings need to be marked or deleted. This makes the methods tedious, time consuming and error prone. 

\begin{figure}[h!]
  \centering
  \includegraphics[width=\linewidth]{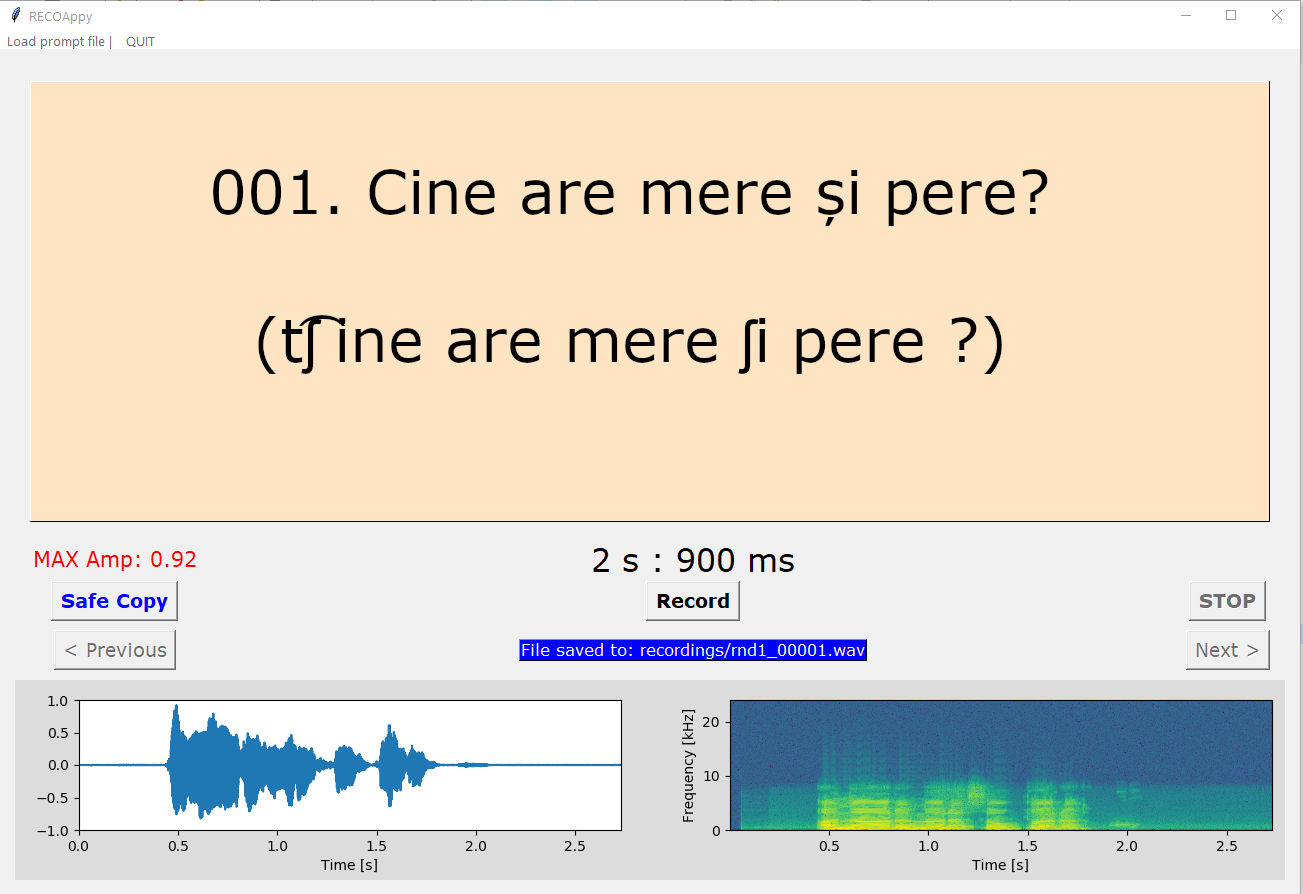}
  \caption{RECOApy GUI}
  \label{fig:recoappy}
  %\vspace{-0.6cm}
\end{figure}

RECOApy was developed with the main objective of streamlining the end-user speech recording process through a series of pre- and post-processing steps. The GUI application is implemented in Python 3.7 with Tkinter\footnote{\url{https://wiki.python.org/moin/TkInter}} and PyAudio.\footnote{\url{https://pypi.org/project/PyAudio/}} Its interface is shown in Figure~\ref{fig:recoappy}. 
Each prompt is individually displayed to the speaker. Once the recording starts, the input amplitude is monitored and its peak value is displayed such that any signal distortion or low level input can be detected. For additional monitoring, the lower panels of the interface display the waveform and spectrogram of the recorded prompt. Parameters such as sampling frequency and bit depth can be set from the configuration file and depend on the available hardware. The recording operator can easily navigate through the prompts and re-record any of them without any extra setup. 
Additional features of RECOApy include waveform normalization and silence trimming, as well as a \emph{Safe Copy} option. This means that if the recording operator is unsure of the correctness of the current recording, a backup copy can be saved and later inspected. 

Alongside the orthographic form of the prompts, the phonetic transcription can also be displayed. This enables the speaker to read the prompts as intended by the developer. The phonetic transcription may already be available in the prompts, or can be generated and saved from within RECOApy, as introduced in the next section.
%The G2P conversion module is described in the next section. 

\vspace{-.2cm}
%%%%%%%%%%%%%%%%%%%%%%%%
%% Phonetic transc
%%%%%%%%%%%%%%%%%%%%%%%%
\section{G2P conversion module}
\label{sec-phon}

To further enhance the usability and applicability of the recording tool, and given the results of \cite{fong2019}, RECOApy can perform an accurate phonetic transcription of the prompts in eight languages: Czech, English, French, German, Italian, Polish, Romanian and Spanish. The data and methods used to develop the grapheme-to-phoneme converters are described next.

%%%%%%%%%%%%%%%%%%%%%%%%
%% Lexicons
%%%%%%%%%%%%%%%%%%%%%%%%
\subsection{Phonetic lexicons}
\label{ssec-lex}

Even for the mainstream languages, large, manually annotated lexicons are not easily and readily accessible. And most research groups have developed their internal resources~\cite{mingzhi2020,45913}. An alternative to this individual effort is the collaborative online resource called Wiktionary. It contains word definitions in 171 languages, of which 45 languages include more than 100,000 entries. The usability of Wiktionary as an alternative to the hand crafted resources has already been studied---\cite{tubiblio104675}~shows its great impact on the future directions of lexicography. A significant number of the dictionary entries also include phonetic transcriptions. Their use in G2P methods has been tackled before \cite{schlippe2014web,deri-knight:2016:ACL}, and can therefore constitute the base for the work presented in this paper.

However, as this resource is constantly expanding, processing the latest database dumps is beneficial \cite{schlippe2014web}.\footnote{\emph{wiktionary-20200301*} versions of the database were used here.} 
A first step for preparing the lexicons was to determine the list of words which include phonetic entries and to extract these pronunciations. Because the data is crowd-controlled, there is no guarantee that the transcriptions are correct and consistent, or that the entries pertain to a single language. To mitigate these issues, a part of the transcriptions were discarded: entries containing graphemes outside the standard alphabet of the respective language; entries containing phonemes whose occurrence is less than 100 across the respective lexicon; and entries with a phonetic transcription significantly longer than the orthographic form, which might be indicative of two or more pronunciation versions entered in the same field. There was also a set of identical entries (same word, same phonetic transcription), and these were collapsed into a single entry.
All lexical stress symbols, if present, were removed.
The final number of entries in each lexicon can be found in Table~\ref{tab-res}.

Due to the potential transcription errors present in Wiktionary, which might affect the performance of the G2P conversion networks, two well-established
manually checked lexicons were also included in the evaluation:
the English CMU Pronunciation Dictionary\footnote{\url{http://www.speech.cs.cmu.edu/cgi-bin/cmudict}} and the Romanian MaRePhor lexicon \cite{toma_sped2017}. Version 0.7b of CMUdict was used and all entries containing numbers and any other symbol except the apostrophe were discarded. The lexical stress in the pronunciation was removed. 

%%%%%%%%%%%%%%%%%%%%%%%%
%% Training
%%%%%%%%%%%%%%%%%%%%%%%%
\subsection{G2P conversion networks}
\label{ssec-net}

Given the variable lengths of the orthographic and phonetic representations of a word, the task of grapheme-to-phoneme conversion is inherently a sequence-to-sequence mapping problem~\cite{SutskeverVL14}. Within the set of sequence-to-sequence deep learning algorithms, the most prominent are those based on recurrent (RNN), convolutional (CNN) and full-attention (Transformer) architectures. Although the RNN seq2seq is a highly efficient and adequate method to process temporal or order dependent sequences, it exhibits a slow convergence and high computational complexity. As a result, more and more NLP tasks have been addressed with CNN or hybrid seq2seq alternatives \cite{GehringAGYD17, DBLP:journals/corr/0001KYS17}. Along the CNN-based architecture, the Transformer network has been successfully applied in machine translation tasks \cite{transformer}, and G2P conversion networks~\cite{mingzhi2020,Yolchuyeva_2019}.

These two seq2seq architectures were selected as the starting point in the development of RECOApy's G2P module. 
The CNN network's encoder and decoder are composed of 1D convolution, activation and normalization layers. An attention layer merges the hidden representations of the encoder and decoder. The attention context is concatenated with the decoder representation and passed through another set of 1D convolution layers---denoted as \emph{decoder output}---to generate a softmax output. No residual connections or embedding layers are used. 
The Transformer network closely follows the architecture of \cite{transformer}, with multi-head self-attention layers combined with fully connected ones in the encoder, decoder and decoder output modules. A positional embedding layer pre-processes the inputs.

For these two neural architectures, the topologies which obtained the best results for English are described in \cite{mingzhi2020,Yolchuyeva_2019,yolchuyeva2}. However, taking into consideration the G2P complexity across languages, as well as the variable dimension of each phonetic lexicon, the architectures' hyperparameters need to be optimized~\cite{MelisDB17}. Genetic algorithms and evolution strategies manage to provide near-optimal solutions for complex tasks, such as image classification \cite{abs-1807-07362} and reinforcement learning \cite{1712-06567}. For the current task of G2P conversion across multiple languages and datasets, an evolution strategy (ES) similar to the one described in \cite{abs-1807-07362} was adopted. The genes represent various topology parameters, such as number of layers in the encoder or the decoder, the hidden dimensions of the layers or the activation function. The fitness of a genome is determined on its ability to predict a set of word-level phonetic transcriptions. The initial population is randomly selected from the genome pool. In each new generation, the fittest individuals are maintained and bred to create new individuals by random recombinations and mutations. A small sample of the less fit individuals are also bred in order to explore the gene space more thoroughly. 

%%%%%%%%%%%%%%%%%%%%%%%%
%% Results
%%%%%%%%%%%%%%%%%%%%%%%%
\vspace{-0.1cm}
\subsection{G2P results}
\label{sec-res}

\begin{table}[t!]
  \caption{Set of genes and gene values used in the evolution strategy. The first column marks the gene ID within the genome.}
  \addtolength{\tabcolsep}{-2pt}
   \renewcommand{\arraystretch}{0.9}
  \label{tab-genes}
  \centering
  \begin{tabular}{l l c  } 
    \toprule
Gene ID&\multicolumn{2}{c }{\textbf{CNN seq2seq }}\\
\midrule
\textbf{G1} & encoder layers& 2, 3, 4\\
\textbf{G2} &encoder layers dimension& 32, 64, 128, 256\\
\textbf{G3} &decoder layers & 2, 3, 4  \\
\textbf{G4} &decoder layers dimension & 32, 64, 128, 256\\
\textbf{G5} &decoder output layers& 2, 3, 4  \\
\textbf{G6} &decoder output layers dim. & 32, 64, 128 \\
\textbf{G7} &activation & ReLU, Linear \\
\textbf{G8} &optimizer & Adam, RMSprop \\
\textbf{G9} &batch size & 32, 64, 128, 256, 512 \\
\midrule
& \multicolumn{2}{c}{ \textbf{Transformer seq2seq}} \\
\midrule
\textbf{G1} &encoder layers & 2, 3, 4 \\
\textbf{G2} &decoder layers & 2, 3, 4 \\
\textbf{G3} &embedding dimension & 32, 64, 128 \\
\textbf{G4} &attention heads & 2, 4 \\
\textbf{G5} &dropout rate & 0.01, 0.05, 0.1, 0.15 \\
\textbf{G6} &hidden layer dimension & 32, 64, 128, 256 \\
&& 512, 1024\\
\textbf{G7} &batch size & 32, 64, 128, 256, 512 \\
  \bottomrule
  \end{tabular}
  \vspace{-0.6cm}
\end{table}

The neural network architectures' hyperparameters were optimized over 10 generations each with a population size of 10. The fitness of a genome was assessed in terms of the word error rate (WER) computed over a held-out test set of 500 samples at the end of a 20 epoch training process. The small number of epochs and evaluation samples was chosen so that the evolution strategy did not fit the respective train-test split. The number of lexicon entries used for hyperparameter optimisation was limited to 150,000 samples.\footnote{See Table~\ref{tab-res} for the number of entries in each lexicon.} The set of genes and gene values for each neural architecture is shown in Table~\ref{tab-genes}. This set does by no means explore the entire hyperparameter search space, but it does address some of the key topological variables.
The fittest individual for each neural architecture, language and lexicon was selected and trained further on the entire set of entries. An early stopping criterion set to monitor variations of less than 1\% in the loss metric over 50 steps prevented overfitting. An 80-20 split with random sampling was employed for training and testing the networks, respectively. The split was different from the one used in the evolution strategy, and the fitness computation data was discarded. 

Table~\ref{tab-res} shows the results of the G2P conversion module. It includes the total number of entries in each lexicon next to the number of unique entries and phonetic symbols. The number of phonetic symbols represent the set of symbols used in the phonetic transcriptions. For the Wiktionary lexicons these might not fully overlap with the language's phoneme set. For each neural architecture the genes of the fittest individual are also presented. The accuracy of the G2P is reported in terms of word error rate (WER) and Levenshtein distance-based phoneme error rate (PER) \cite{levenshtein1966bcc}. For entries with multiple pronunciations, the target which minimized the PER and WER was selected. 

%\textcolor{red}{Update the GENES!!!!}

\begin{table*}[th!]
\renewcommand{\arraystretch}{0.96}
  \caption{Lexicon descriptions, network hyperparameters and accuracy results of the grapheme-to-phoneme module . The phonetic symbols column indicates the number of distinct phonemes found in the respective lexicon. The gene IDs are listed in Table~\ref{tab-genes}. Best results for each lexicon are highlighted in boldface.}
  \addtolength{\tabcolsep}{-4pt}
  \label{tab-res}
  \centering
  \begin{tabular}{l l c c c c c c c c c c c c c | c c}
    \toprule
   \multirow{2}{*}{\textbf{Lang}}& \multirow{2}{*}{\textbf{Lexicon}}& \multirow{2}{*}{\textbf{Entries}}& \textbf{Unique} & \textbf{Phonetic} &\multirow{2}{*}{\textbf{Model}}&\multirow{2}{*}{G1}  & \multirow{2}{*}{G2}& \multirow{2}{*}{G3} & \multirow{2}{*}{G4} & \multirow{2}{*}{G5} & \multirow{2}{*}{G6} &\multirow{2}{*}{G7}& \multirow{2}{*}{G8}&\multirow{2}{*}{G9}&\multirow{2}{*}{ \textbf{WER}}& \multirow{2}{*}{\textbf{PER}} \\
    & & &\textbf{entries} & \textbf{symbols}&&&&&&&&&&&&   \\
  
    %%english
    \midrule
    \multirow{4}{*}{\textbf{EN}} &  \multirow{2}{*}{\textbf{CMUdict}} & \multirow{2}{*}{132,585}  & \multirow{2}{*}{123,874} & \multirow{2}{*}{39} 
    					            &CNN & 2 &128 &2 &128 &3 &128/64/32&ReLU &RMSp& 512 & 29.82&11.41 \\
    					     &&&&& Transformer&4&3&64&4&0.01&512&64&-&-&\textbf{23.16}&\textbf{8.03}\\
    					     %wikt

     					     \cline{2-17} \noalign{\vskip 1mm}    
     					     
     					      & \multirow{2}{*}{Wiktionary} &  \multirow{2}{*}{71,332}&  \multirow{2}{*}{48,773}  & \multirow{2}{*}{39} 
    						      &CNN&2&128&2&128&3 &128/64/32&ReLU &RMSp& 256& 28.92&12.39\\
     					     &&&&& Transformer&4&4&32&4&0.01&128&128&-&-&\textbf{22.50}&\textbf{8.23}\\				
    					     
   %% romanian				        				
    \midrule
    \multirow{4}{*}{\textbf{RO}} &\multirow{2}{*}{\textbf{MaRePhor}} & \multirow{2}{*}{72,375}  & \multirow{2}{*}{72,375} &  \multirow{2}{*}{40} 
    					        &CNN& 3&64&2&32&3&64/32/32&Lin &Adam&128&2.64&0.5\\
    					     &&&&& Transformer&2&4&32&2&0.05&64&64&-&-&\textbf{2.30}&\textbf{0.42} \\
    
     					     %%wikt	
     					     \cline{2-17} \noalign{\vskip 1mm}   
     					     & \multirow{2}{*}{Wiktionary} &  \multirow{2}{*}{63,013}&  \multirow{2}{*}{ 62,733}  & \multirow{2}{*}{32} 
    						&CNN& 3&128&2&32&3&128/64/32&Lin&Adam&512&\textbf{3.00}&\textbf{0.50}\\
     					     &&&&& Transformer&3&2&64&2&0.05&64&256&-&-&3.58&0.71\\

     \midrule

    %czech  
     \multirow{2}{*}{\textbf{CZ}} & \multirow{2}{*}{Wiktionary} & \multirow{2}{*}{42,014} & \multirow{2}{*}{41,419} & \multirow{2}{*}{41}
     						&CNN&2&32&4&128&3&64/32/32&Lin&RMSp&128&11.69&3.84\\
    						&&&&& Transformer&2&2&32&2&0.05&64&32&-&-&\textbf{9.45}&\textbf{2.37} \\    					     
    %%german
    \midrule
     \multirow{2}{*}{\textbf{DE}} & \multirow{2}{*}{Wiktionary} & \multirow{2}{*}{327,296} & \multirow{2}{*}{315,793} & \multirow{2}{*}{51}
     						&CNN&3&128&3&32&3&128/64/32&ReLU&Adam&512&\textbf{5.50}&\textbf{1.43}\\
    						&&&&& Transformer&4&2&64&2&0.05&32&64&-&-&8.80&2.24 \\    				
    						
       %spanish
     \midrule
    \multirow{2}{*}{\textbf{ES} }& \multirow{2}{*}{Wiktionary} & \multirow{2}{*}{ 49,346}& \multirow{2}{*}{42,732} & \multirow{2}{*}{31}
    						&CNN&3&128& 4&64&2&128/64&ReLU&Adam&128&\textbf{9.81}&\textbf{2.20}\\
    						&&&&& Transformer&2&4&32&4&0.05&32&32&-&-&11.90&2.95 \\    
    											
    %% french
    \midrule
     \multirow{2}{*}{\textbf{FR}}&  \multirow{2}{*}{Wiktionary} &  \multirow{2}{*}{1,121,714}& \multirow{2}{*}{1,115,343} & \multirow{2}{*}{35}
     						&CNN&3&128&3&32&3&128/64/32&ReLU&Adam&512&\textbf{4.38}&1.02\\	
     						&&&&& Transformer&2&3&64&2&0.05&128&64&-&-&4.78&\textbf{0.97}\\

    %% italian					
    \midrule
     \multirow{2}{*}{\textbf{IT}} & \multirow{2}{*}{Wiktionary} &  \multirow{2}{*}{29,826} &  \multirow{2}{*}{29,242}& \multirow{2}{*}{28}
     						&CNN&2&128&4&128&2&64/32&ReLU&RMSp&256&\textbf{18.67}&\textbf{4.44}\\
    						&&&&& Transformer& 2&2&64&2&0.01&512&64&-&-&19.04&5.00 \\

    %polish
     \midrule
     \multirow{2}{*}{\textbf{PL} }& \multirow{2}{*}{Wiktionary} & \multirow{2}{*}{35,646} &  \multirow{2}{*}{35,544} &  \multirow{2}{*}{48}
     						&CNN&4&64&2&128&2&128/64&ReLU&Adam&128&3.59&1.84\\
    						&&&&& Transformer&3&2&64&4&0.05&1024&128&-&-&\textbf{2.98}&\textbf{1.34} \\

  %  encoder_nume, decoder_num, embedding, attention heads & dropout rate & hidden_dim & batch_size
 
    \bottomrule
  \end{tabular}
  \vspace{-0.2cm}
\end{table*}

The best performing architecture varies across languages, as well as in between lexicons of the same language, but the error rate differences are not truly significant. For example, the Romanian Wiktionary lexicon is better fitted by the CNN seq2seq, while for MaRePhor, the Transformer achieves lower WER and PER. For English, both lexicons are better fitted by the Transformer. 
The dataset's dimension does not seem to favour any of the architectures either, even though the number of trainable parameters is largely different. For example, the MaRePhor CNN model has 173,672 trainable parameters, and the transformer has only 71,146. But by inspecting the comparable sized lexicons in Czech and Spanish, the Transformer achieves better WER and PER for Czech, yet falls short of the CNN seq2seq in Spanish. This happens despite the fact that Czech and Spanish also exhibit comparable orthographic transparency levels \cite{gillon}.
One conclusion that can be directly drawn from here is that there is no universal recipe to solve the G2P task, and each solution and architecture needs to be tailored to the particular language, phonetic representation, and available resources. 
The absolute error rates for each language presented here are comparable or lower than the ones in \cite{schlippe2014web} and \cite{45913}. But the different lexicon versions and train-test splits make a direct, fully correct comparison impossible. As an overview of the architectures' performance, the average WER across lexicons for the CNN seq2seq is 11.80\%, and the PER is 3.95\%. For the Transformer, the average WER is 10.95\% and PER is 3.22\%.

Inspecting the performance over the supervised lexicons, for MaRephor the results are in line with previous studies~\cite{stan_sped2019}. The CMUdict error rates obtained here (23.16\% WER and 8.03\% PER) are slightly lower than the ones reported in the state-of-the-art methods (\cite{Yolchuyeva_2019}:~22.1\% WER and 5.1\% PER). However, the CMUdict versions and train-test splits are different. When applying the same architecture\footnote{The authors of~\cite{Yolchuyeva_2019} kindly provided their implementation.} on this version of the CMUdict, the results were 22.8\% WER and 7.19\% PER. 
It is interesting, however, to notice that the ES evolved a rather similar architecture for the Transformer seq2seq. It may be the case that an evaluation of the fitness over larger number of epochs and validation set, would yield the same architecture, and therefore same performance. 
One other interesting fact in the results reported here is the high WER for Italian. When analysing the decoded sequences from both networks, it was found that over 60\% of the erroneous words had only a single incorrect phoneme, and it was mostly the case of vowel-semivowel substitutions. 

Looking at the inference duration, the MaRePhor CNN seq2seq model processes 5000 words in approximately 55 seconds, while the Transformer seq2seq does it in around 120 seconds.\footnote{On an NVidia GeForce RTX 2080 Ti GPU with 12GB vRAM.}
Given the large difference in inference time and only minor drops of accuracy for some of the lexicons, RECOApy integrates the CNN-based models alone. However, the trained Transformer models are available in the tool's webpage.

 %at the training time for the two architectures, one epoch of CNN seq2seq takes on average 8 seconds, for the CMUdict architecture. 

%The only requirement is that the input text is normalized, as the module discards any non-alphabetic symbols.

%%%%%%%%%%%%%%%%%%%%%%%%
%% Conclusions
%%%%%%%%%%%%%%%%%%%%%%%%
\vspace{-0.2cm}
\section{Conclusions}
\label{sec-conc}

This paper introduced RECOApy, a tool for data recording, pre-processing and phonetic transcription of training data aimed at
speech-based end-to-end applications. The tool enables fast and accurate recording of text prompts at various sampling rates and bit depths, while offering the recording operator the possibility to supervise the quality of the process as well. Additional automatic options to normalise the audio and to discard the start and end silence segments are also available. One other important feature of RECOApy is that of automatic phonetic transcription of the prompts in eight languages: Czech, English, French, German, Italian, Polish, Romanian and Spanish. The G2P module consists of state-of-the-art neural network based architectures achieving low word and phoneme error rates across all languages. As a conclusion, the RECOApy tool can most certainly be used as a reliable means to develop the training data for end-to-end speech-based applications. In fact, our research group has already collected over 50 hours of prompted speech from non-expert volunteers using this recording tool. The tool, lexicons and models are available here: \url{https://gitlab.utcluj.ro/sadriana/recoapy/}.

Future developments of the tool include the addition of more languages in the G2P module, a more in-depth analysis of the hyperparameter space, as well as the augmentation of the prompts with syllabification and lexical stress assignment. A potential significant developement would be to also include prosodic cues---similar to~\cite{WilhelmsTricarico2013TheLT}.

\vspace{-0.3cm}
\section{Acknowledgement}
This work was funded through a grant from the Romanian Ministry of Research and Innovation, PCCDI – UEFISCDI, project number PN-III-P1-1.2-PCCDI-2017-0818/73.

%This work was supported by a grant of the Romanian Ministry of Research and Innovation, PCCDI – UEFISCDI, project number PN-III-P1-1.2-PCCDI-2017-0818/73, within PNCDI III. 

%\clearpage
\bibliographystyle{IEEEtran}

\bibliography{master}

\begin{thebibliography}{10}
\providecommand{\url}[1]{#1}
\csname url@samestyle\endcsname
\providecommand{\newblock}{\relax}
\providecommand{\bibinfo}[2]{#2}
\providecommand{\BIBentrySTDinterwordspacing}{\spaceskip=0pt\relax}
\providecommand{\BIBentryALTinterwordstretchfactor}{4}
\providecommand{\BIBentryALTinterwordspacing}{\spaceskip=\fontdimen2\font plus
\BIBentryALTinterwordstretchfactor\fontdimen3\font minus
  \fontdimen4\font\relax}
\providecommand{\BIBforeignlanguage}[2]{{%
\expandafter\ifx\csname l@#1\endcsname\relax
\typeout{** WARNING: IEEEtran.bst: No hyphenation pattern has been}%
\typeout{** loaded for the language `#1'. Using the pattern for}%
\typeout{** the default language instead.}%
\else
\language=\csname l@#1\endcsname
\fi
#2}}
\providecommand{\BIBdecl}{\relax}
\BIBdecl

\bibitem{DeepSpeech}
A.~Y. Hannun, C.~Case, J.~Casper, B.~Catanzaro, G.~Diamos, E.~Elsen,
  R.~Prenger, S.~Satheesh, S.~Sengupta, A.~Coates, and A.~Y. Ng, ``Deep speech:
  Scaling up end-to-end speech recognition,'' \emph{ArXiv}, vol. abs/1412.5567,
  2014.

\bibitem{Tacotron2}
J.~Shen, R.~Pang, R.~J. Weiss, M.~Schuster, N.~Jaitly, Z.~Yang, Z.~Chen,
  Y.~Zhang, Y.~Wang, R.~J. Skerry{-}Ryan, R.~A. Saurous, Y.~Agiomyrgiannakis,
  and Y.~Wu, ``Natural {TTS} synthesis by conditioning wavenet on mel
  spectrogram predictions,'' in \emph{Proceedings of ICASSP}, 2018.

\bibitem{DeepVoice}
S.~{\"{O}}. Arik, M.~Chrzanowski, A.~Coates, G.~Diamos, A.~Gibiansky, Y.~Kang,
  X.~Li, J.~Miller, J.~Raiman, S.~Sengupta, and M.~Shoeybi, ``{Deep Voice:
  Real-time Neural Text-to-Speech},'' in \emph{Proceedings of ICML}, 2017.

\bibitem{fong2019}
J.~Fong, J.~Taylor, and K.~Richmond, ``{A Comparison of Letters and Phones as
  Input to Sequence-to-Sequence Models for Speech Synthesis},'' in \emph{Proc.
  of Interspeech}, 2019, pp. 223--227.

\bibitem{48239}
V.~Wan, C.~an~Chan, T.~Kenter, J.~Vit, and R.~Clark, ``Chive: Varying prosody
  in speech synthesis with a linguistically driven dynamic hierarchical
  conditional variational network,'' in \emph{Proceedings of the 36th
  International Conference on Machine Learning (ICML 2019)}, 2019, pp.
  3331--3340.

\bibitem{gillon}
G.~Gillon, \emph{{Phonological Awareness 2nd Edition: From Research to
  Practice}}.\hskip 1em plus 0.5em minus 0.4em\relax The Guilford Press, 2018.

\bibitem{schlippe2014web}
T.~Schlippe, S.~Ochs, and T.~Schultz, ``Web-based tools and methods for rapid
  pronunciation dictionary creation,'' \emph{Speech Communication, 118, January
  2014.}, vol.~56, p. 101, 2014.

\bibitem{deri-knight:2016:ACL}
A.~Deri and K.~Knight, ``Grapheme-to-phoneme models for (almost) any
  language,'' in \emph{Proceedings of the 2016 Conference of the Association
  for Computational Linguistics}.\hskip 1em plus 0.5em minus 0.4em\relax
  Berlin, Germany: Association for Computational Linguistics, August 2016.

\bibitem{mingzhi2020}
M.~Yu, H.~Nguyen, A.~Sokolov, J.~Lepird, K.~Sathyendra, S.~Choudhary,
  A.~Mouchtaris, and S.~Kunzmann, ``{Multilingual Grapheme-to-Phoneme
  Conversion with Byte Representation},'' in \emph{Proc. of ICASSP}, 2020.

\bibitem{route-etal-2019-multimodal}
J.~Route, S.~Hillis, I.~Czeresnia~Etinger, H.~Zhang, and A.~W. Black,
  ``{Multimodal, Multilingual Grapheme-to-Phoneme Conversion for Low-Resource
  Languages},'' in \emph{Proceedings of the 2nd Workshop on Deep Learning
  Approaches for Low-Resource NLP (DeepLo 2019)}.\hskip 1em plus 0.5em minus
  0.4em\relax Hong Kong, China: Association for Computational Linguistics, Nov.
  2019, pp. 192--201.

\bibitem{Yao2015SequencetosequenceNN}
K.~Yao and G.~Zweig, ``Sequence-to-sequence neural net models for
  grapheme-to-phoneme conversion,'' in \emph{Proc. of Interspeech}, 2015.

\bibitem{43264}
K.~Rao, F.~Peng, H.~Sak, and F.~Beaufays, ``Grapheme-to-phoneme conversion
  using long short-term memory recurrent neural networks,'' in \emph{ICASSP},
  2015.

\bibitem{Milde2017MultitaskSM}
B.~Milde, C.~Schmidt, and J.~K{\"o}hler, ``{Multitask Sequence-to-Sequence
  Models for Grapheme-to-Phoneme Conversion},'' in \emph{Proc. of Interspeech},
  2017.

\bibitem{45913}
D.~van Esch, M.~Chua, and K.~Rao, ``Predicting pronunciations with
  syllabification and stress with recurrent neural networks,'' in
  \emph{Proceedings of Interspeech}, 2016.

\bibitem{Juzov2019UnifiedLD}
M.~Juzov{\'a}, D.~Tihelka, and J.~Vit, ``{Unified Language-Independent
  DNN-Based G2P Converter},'' in \emph{Proceedings of Interspeech}, 2019.

\bibitem{conf/interspeech/SjolanderB00}
K.~Sjölander and J.~Beskow, ``Wavesurfer - an open source speech tool.'' in
  \emph{Proceedings of Interspeech}.\hskip 1em plus 0.5em minus 0.4em\relax
  ISCA, 2000, pp. 464--467.

\bibitem{tubiblio104675}
C.~M. Meyer and I.~Gurevych, ``Wiktionary: A new rival for expert-built
  lexicons? exploring the possibilities of collaborative lexicography,'' in
  \emph{Electronic Lexicography}, S.~Granger and M.~Paquot, Eds.\hskip 1em plus
  0.5em minus 0.4em\relax Oxford: Oxford University Press, November 2012, pp.
  259--291.

\bibitem{toma_sped2017}
S.-A. Toma, A.~Stan, M.-L. Pura, and T.~Barsan, ``{MaRePhoR - An Open Access
  Machine-Readable Phonetic Dictionary for Romanian},'' in \emph{{Proceedings
  of the 9th Conference on Speech Technology and Human-Computer Dialogue
  (SpeD)}}, Bucharest, Romania, July, 6-9 2017.

\bibitem{SutskeverVL14}
I.~Sutskever, O.~Vinyals, and Q.~V. Le, ``{Sequence to Sequence Learning with
  Neural Networks},'' in \emph{Proceedings of the 27th International Conference
  on Neural Information Processing Systems - Volume 2}, ser. NIPS’14, 2014,
  p. 3104–3112.

\bibitem{GehringAGYD17}
J.~Gehring, M.~Auli, D.~Grangier, D.~Yarats, and Y.~N. Dauphin,
  ``{Convolutional Sequence to Sequence Learning},'' in \emph{Proceedings of
  the 34th International Conference on Machine Learning}, ser. Proceedings of
  Machine Learning Research, vol.~70, 2017, pp. 1243--1252.

\bibitem{DBLP:journals/corr/0001KYS17}
W.~Yin, K.~Kann, M.~Yu, and H.~Sch{\"{u}}tze, ``Comparative study of {CNN} and
  {RNN} for natural language processing,'' \emph{CoRR}, vol. abs/1702.01923,
  2017.

\bibitem{transformer}
A.~Vaswani, N.~Shazeer, N.~Parmar, J.~Uszkoreit, L.~Jones, A.~N. Gomez, L.~u.
  Kaiser, and I.~Polosukhin, ``Attention is all you need,'' in \emph{Advances
  in Neural Information Processing Systems 30}, I.~Guyon, U.~V. Luxburg,
  S.~Bengio, H.~Wallach, R.~Fergus, S.~Vishwanathan, and R.~Garnett, Eds.\hskip
  1em plus 0.5em minus 0.4em\relax Curran Associates, Inc., 2017, pp.
  5998--6008.

\bibitem{Yolchuyeva_2019}
S.~Yolchuyeva, G.~Németh, and B.~Gyires-Tóth, ``Transformer based
  grapheme-to-phoneme conversion,'' \emph{Proceedings of Interspeech}, Sep
  2019.

\bibitem{yolchuyeva2}
------, ``{Grapheme-to-Phoneme Conversion with Convolutional Neural
  Networks},'' \emph{Applied Sciences}, vol.~9, no.~6, p. 1143, 2019.

\bibitem{MelisDB17}
G.~Melis, C.~Dyer, and P.~Blunsom, ``On the state of the art of evaluation in
  neural language models,'' \emph{CoRR}, vol. abs/1707.05589, 2017.

\bibitem{abs-1807-07362}
T.~Hinz, N.~Navarro, S.~Magg, and S.~Wermter, ``Speeding up the hyperparameter
  optimization of deep convolutional neural networks,'' \emph{International
  Journal of Computational Intelligence and Applications}, vol.~17, pp.
  1\,850\,008:1--1\,850\,008:15, 2018.

\bibitem{1712-06567}
F.~P. Such, V.~Madhavan, E.~Conti, J.~Lehman, K.~O. Stanley, and J.~Clune,
  ``{Deep Neuroevolution: Genetic Algorithms Are a Competitive Alternative for
  Training Deep Neural Networks for Reinforcement Learning},'' \emph{CoRR},
  vol. abs/1712.06567, 2017.

\bibitem{levenshtein1966bcc}
V.~Levenshtein, ``{Binary Codes Capable of Correcting Deletions, Insertions and
  Reversals},'' \emph{Soviet Physics Doklady}, vol.~10, p. 707, 1966.

\bibitem{stan_sped2019}
A.~Stan, ``{Input Encoding for Sequence-to-Sequence Learning of Romanian
  Grapheme-to-Phoneme Conversion},'' in \emph{{Proceedings of the 10th IEEE
  International Conference on Speech Technology and Human-Computer Dialogue
  (SpeD)}}, Timisoara, Romania, October, 10-12 2019.

\bibitem{WilhelmsTricarico2013TheLT}
R.~Wilhelms-Tricarico, J.~B. Reichenbach, and G.~Marple, ``{The Lessac
  Technologies Hybrid Concatenated System for Blizzard Challenge 2013},'' in
  \emph{Proceedings of Blizzard Challenge}, 2013.

\end{thebibliography}

\end{document}